\documentclass{aastex}

\usepackage{graphicx,shapepar}
\newcommand{\kms}{\ensuremath{\mathrm{km\ s}^{-1}}}
\newcommand{\grados}{$\,^\mathrm{o}\,$}
\newcommand{\vhel}{\ensuremath{V_\odot}}
\newcommand\SIIlam{[\ion{S}{2}]\,6731\,\AA\@}

\newcommand\NIIlam{[\ion{N}{2}]\,6584\,\AA\@}
\newcommand\OIIIlam{[\ion{O}{3}]\,5007\,\AA\@}

\newcommand\Ha{\ensuremath{\mathrm{H}\alpha}}

\newcommand\SII{[\ion{S}{2}]}
\newcommand\OIII{[\ion{O}{3}]}
\newcommand\NII{[\ion{N}{2}]}

\title{A spectroscopic and photometric study of the planetary nebulae Kn~61 and Pa~5}

\author{Ma.~T. Garc\'{\i}a-D\'{\i}az\altaffilmark{1}, D.  Gonz\'alez-Buitrago\altaffilmark{1},
  J. A. L\'opez\altaffilmark{1}, S. Zharikov\altaffilmark{1}, G. Tovmassian\altaffilmark{1}, N. Borisov\altaffilmark{2}, G. Valyavin \altaffilmark{2}} 
  \altaffiltext{1}{Instituto de
    Astronom\'ia, Universidad Na\-cio\-nal Aut\'o\-no\-ma de
    M\'e\-xico. Km 103 Carretera Tijuana-Ensenada, 22860 Ensenada, Baja California,
    M\'exico} 
  \altaffiltext{2}{Special Astrophysical Observatory of the RAS,
    369167, Nizhny Arkhyz, Karachaevo-Cherkesia, Russia}
\email{tere, dgonzalez, jal, zhar,
  gag@astro.unam.mx, borisov@sao.ru, gvalyavin@gmail.com}

\begin{abstract}

We present the first morpho-kinematical analysis of  the planetary
nebulae Kn~61 and Pa~5 and explore the nature of their central stars.
Our analysis is based on high resolution and
medium resolution spectroscopic observations, deep
narrow-band imaging, and integral photometry. This material allows us
to identify the morphological components and study their
kinematics. The direct images and spectra indicate an absence of the
characteristic \NII{} and \SII{} emission lines in both nebulae.  The
nebular spectrum of Kn~61 suggests an hydrogen
deficient planetary nebula and the stellar spectrum of the central
star reveals a hydrogen deficient PG~1159-type star. The \OIII\,
position velocity diagram reveals that Kn~61 is a closed, empty,
spherical shell with a thin border and a filamentary surface expanding
at 67.6 \kms\, and the shell is currently not expanding isotropically.
We derived a kinematic age of $\sim$~1.6$\times$10$^4$~yrs for an
assumed distance of 4~kpc.  A photometric period of
$\sim$ 5.7($\pm$0.4) days has been detected for Kn 61, indicating 
presence of a possible binary system at its core.  A possible link
between filamentary, spherical shells and PG~1159-type stars is noted.

The morphology of Pa~5 is dominated by an equatorial toroid and faint
polar extensions. The equatorial region of this planetary nebula is
expanding at 45.2~\kms.  The stellar spectrum corresponds to a very
hot star and is dominated by a steep blue rising continuum and He~II,
Balmer and Ca~II photospheric lines.

\end{abstract}


\begin{document}
 \maketitle

\section{Introduction}
\label{sec:introduction}

In the present study we analyze the characteristics of two planetary
nebulae, namely, Kn~61 which has a nearly perfect spherical shape and
Pa~5, which presents a toroidal structure with faint polar extensions.
We also explore the possibility that these PNe host binary cores, as
has been suggested by Long et al. (2013), Kronberger et al. (2011),
Douchin et al. (2011) and Jacoby et al. (2010).

The PN Kn~61 (nicknamed the Soccer Ball Nebula) was discovered
by Kronberger et al. (2011). A high quality optical image in the
combined light of \OIIIlam\, and \Ha\, has been published by the
GEMINI observatory (http://www.gemini.edu/node/11656).  Of particular
interest is that this object lies within the Kepler field. Long et
al. (2013) report brightness  variability with a
  possible periodicity up to $\sim$ 6 days from Kepler data. The
composite \OIII\, and H$\alpha$ Gemini image (see Figure 1 from
Douchin et al. 2011) reveals a predominantly bright \OIII\,
filamentary, spherical, shell. The central star of Kn~61, also known
as SDSS J192138.93+381857.2 has magnitude $u$ = 18.66($\pm$0.01), $g$
= 18.03($\pm$0.01), $r$ = 18.48($\pm$0.01).  The (uncertain) distance
has been assumed to be 4~kpc (http://www.gemini.edu/node/11656), from
the GEMINI note on the object, mentioned above, though given its
angular size of nearly 110 arcsec it is possible that it could be
slightly closer.

Pa~5 was discovered by D. Patchick (Jacoby et al. 2010). This nebula
also lies within the Kepler field. {\O}stensen, et al. (2010) using
low resolution spectroscopy in the spectral range 3500~--~5300~\AA,
found that the central star of Pa~5, J19195$+$4445, is a very hot
one. They identify C~IV and He~II lines in the 4640 - 4686\,\AA\,
region in the stellar spectrum as indicative of the PG~1159 class.  In
addition, they find a period of 1.12 days using the Kepler data.  The
origin of variability of the central star ($\approx$0.002 mag) is
reported as unknown. Kronberger et al. (2006) have classified Pa~5 as
a probable bipolar PN. In direct images (DSS) it looks like a bipolar
but only the northeast lobe is apparent with the opposite lobe being
much fainter.
 
In order to improve our understanding of these two PNe contained
within the Kepler field we present new spectroscopic and CCD imaging
and photometric results for the nebulae and their central stars. This
work combines results from datasets obtained at the Observatorio
Astron\'omico Nacional at San Pedro M\'artir, B. C., M\'exico
(OAN-SPM) and from the Special Astrophysical Observatory 6-m telescope
(SAO-RAS).

The structure of the paper is as follows. In \S\ref{sec:observations}
we describe the observations and the data-reduction steps. In
\S\ref{sec:Results} we discuss the results. In \S\ref{sec:conclusion}
we summarize the conclusion of this work.

\section{The Observations}
\label{sec:observations}

Tables~1 and 2 display the information relative to the observing runs,
divided into instruments with respective dates, exposure times, number
of spectra and spectral range. All data were reduced using standard
procedures in IRAF\footnote{IRAF is distributed by the National
  Optical Astronomy Observatories, which is operated by the
  Association of Universities for Research in Astronomy, Inc. under
  cooperative agreement with the National Science foundation} to
correct bias, cosmic rays extract and wavelength calibrate the one and
two-dimensional spectra.


\subsection{Optical Imaging and Photometry}

The monochromatic images of Kn~61 and Pa~5 were obtained on 2012,
August 3 for Kn~61, and on 2013, August 9 for Pa~5, using the Mexman
CCD photometer on the 84~cm telescope at the OAN-SPM. We used the
ESOPO CCD detector with 2048 $\times$ 4612 square pixels, 15 $\mu$m
(0.25$\times$0.25 arcsec) pixel size.  Kn~61 and Pa~5 were observed in
the light of \OIIIlam, \Ha, \SIIlam\, and \NIIlam\, with bandwidths
of 52~\AA, 11~\AA, 54~\AA\, and 10~\AA, respectively. The exposure
time for each filter was 1800~s. The data were corrected for bias,
cosmic rays hits and flat-fielded in order to remove the
pixel-to-pixel response.

Figures 1 and 2 are mosaics of images of Kn~61 and Pa~5,
respectively. For each figure, the panels correspond to \OIII\, (left
panel), and H$\alpha$ (right panel). We did not show the \NII\, nor
\SII\, images because the nebulae do not have any emission in these
lines.

Kn~61 and Pa~5 were also observed with the 1.5~m telescope at the OAN-SPM,
and the RATIR camera (Butler et al. 2012), in the $r$ and $i$ filters
simultaneously, for 18 consecutive days with half hour observations
per night per object. The exposure time was 90 seconds for each
filter.

We performed differential photometry using the IRAF package apphot
selecting comparison stars in the field for Kn~61 and Pa~5.

\subsection{Spectroscopy Observations}
\label{sec:spectroscopy}

\subsubsection{High resolution}

Long-slit high-resolution spectroscopic observations of the nebulae
Kn~61 and Pa~5 were obtained at the OAN-SPM, M\'exico, on 2012, May
25, and are available in {\it ``The San Pedro M\'artir Kinematic
  Catalogue of Galactic Planetary Nebulae''} (L\'opez et al. 2012).

The data were obtained using the Manchester Echelle Spectrometer,
MES-SPM, (Meaburn et al. 2003) on the 2.1 m telescope in its $f$/7.5
configuration The MES-SPM is a long-slit, echelle spectrometer that
has no cross-disperser; it isolates single orders using interference
filters. For the present observations we used two filters of 90~\AA\,
and 50~\AA\, bandwidths to isolate the 87th and 114th orders
containing the H$\alpha$ $+$ \NIIlam, and \OIIIlam\, nebular emision
lines, respectively.  This instrument was equipped with a Marconi CCD
detector with 2048\,$\times$\,2048 square pixels, each 13.6 $\mu$m on
a side.  The detector was set to a binning of 3\,$\times$\,3 in both
the spatial and spectral directions. 

We used a 150~$\mu$m\, slit of 1\farcs9 and 5\farcm47 projected width
and length, respectively. The resultant spectral dispersion was
approximately 0.26 \AA\,pixel$^{-1}$ and 0.19~\AA\,pixel$^{-1}$ at
H$\alpha$ and \OIIIlam, respectively (equivalent to $\sim$ 11.5 \kms).
For Kn~61 we placed the slit at a position angle, P.A. = 0\grados and
through the central star of the nebula (see Figure~1). For Pa~5 we
placed the slit along the bright elongated structure whose axis is
oriented at P.A.~=~50\grados (see Figure~2). A second exposure along
the axis defined by the lobes was obtained but it turned out of poor
signal to noise due to thick clouds.  Images of the slit projected on
the sky were obtained to establish the exact position of the slit on
the target.  Wavelength calibration against the spectrum of a Th/Ar
lamp was obtained after each science exposure, this yields an accuracy
of $\pm$1 \kms\, when converted to radial velocity.  Spectra presented
in this work are corrected to heliocentric velocity (\vhel).  The
bi-dimensional H$\alpha$ and \OIIIlam\, emission line spectra or
position--velocity (P--V) arrays are shown in Figures~3 and 4 for Kn~61
and Pa~5, respectively.

\subsubsection{Medium resolution}
\subsubsubsection{OAN-SPM}

Medium resolution spectroscopy was performed with the 2.1~m telescope
at the OAN-SPM in 2012 and 2013.  We used the Boller \& Chivens
(B\&Ch) spectrograph using a 600 l mm$^{-1}$ grating with a dispersion
of 1.17~\AA\,pixel$^{-1}$. The spectra span the wavelength range from
4000~\AA\, to 7000~\AA.  We also used a 1200 l mm$^{-1}$ grating with
a dispersion 0.60 \AA\, pixel$^{-1}$ covering the wavelength range
from 4560~\AA\, to 5750~\AA.  A Marconi CCD with 2048 $\times$ 2048
pixels and no binning was used. In all observations the slit was
oriented east-west and on the central star for each planetary
nebula. The exposure time was 1200 s per spectrum.  A He-Ne-Ar
comparison lamp was obtained after every third spectra to ensure good
wavelength calibration. For flux calibration a standard star was
observed at the end of the night, taken from the list of Landolt
(1992) and Bohlin et al. (2001).

Figures~5 y 6 are mosaics of the stellar and nebular spectra for Kn~61
and Pa~5 respectively, taken from OAN-SPM. The lefthand panels show
medium resolution spectra of the central star, while the righthand
panels show the nebular spectra. Panels a) and b) were taken with the
600 l mm$^{-1}$ grating, while the lower panels, panels c) and d),
were taken with the 1200 l mm$^{-1}$ grating. The 1200 l mm$^{-1}$
grating spectra shown in these figures were averaged over a single
night of observation to improve the signal-to-noise. The spectra for
the nebulae were extracted from the full length of the slit that covers
the nebulae.

\subsubsubsection{SAO-RAS}

Medium-resolution spectroscopic data of Kn~61 were recently obtained
at the Special Astrophysical Observatory 6-m telescope on 2013,
September 12, using the SCORPIO spectrograph (Afanasiev \& Moiseev
2005) installed at the prime focus of the SAO 6-m telescope of the
Russian Academy of Sciences in the long-slit unit mode. In this mode,
the slit dimensions are 6\arcmin\, long $\times$ 1\arcsec\, wide. We
used the CCD detector EEV42-40 and the grism VPHG 1200 g (1200 lines
mm$^{-1}$ with a spectral dispersion of 0.88 \AA\,
pixel$^{-1}$). Three spectra were taken in the spectral range between
3900~\AA\, and 5700~\AA. The standard star (BD28d4211) was observed
for flux calibration. The calibrated spectra are shown in Figure~7.

\section{Results and Discussion}
\label{sec:Results}

\subsection{Kn 61}

Figure~1 shows the images of Kn~61 in \OIII, and  H$\alpha$. We
did not detect any emission in \NII\,6584~\AA\, nor in
\SII\,6717,6731~\AA\, in our images. The \OIII\, image shows a highly
filamentary bubble. The H$\alpha$ image is extremely faint showing
some limb brightening at the border. The outline of the bubble in the
\OIII\, and H$\alpha$ images cover the same area.  In other
hydrogen-deficient nebulae, such as Abell~30 (e.g. Meaburn \& L\'opez
1996) and Abell~78 (e.g. Meaburn et al. 1998), the faint hydrogen
shell clearly extends beyond the \OIII\, shell indicating that the
latter was expelled in a late thermal flash once the hydrogen layer of
the central star had been consumed and the hydrogen-rich envelope had
already expanded. In the case of Kn~61 the nebular emission points
towards an already hydrogen impoverished shell as it left the AGB.

The echelle, long-slit, spectra of the nebular shell of Kn 61 (see
Figure~3) confirm the same peculiar line emission ratios in the sense
that the \Ha\, emission line is extremely faint and the \NII\, lines
are absent whereas \OIII\, is bright.  The line profiles (see Figure
3) indicate that Kn~61 is a closed, hollow, spherical shell with a
thin border.  The filamentary surface is revealed in the \OIII\, line
profile by the knotty structure along the thin border of the profile.
From the \OIII\, P--V array we find that the velocity splitting at the
center of the profile amounts to 135.3 \kms\, with the blueshift
component at \vhel~=~$-93.0$~\kms\, and the redshift component at
\vhel~=~$+$42.3~\kms. This yields an expansion velocity of 67.6~\kms,
which is high for PNe like this (Pereyra et al. 2013). The systemic
heliocentric velocity, as measured from the midpoint between the line
splitting is -25.4($\pm$2)~\kms.  A departure from perfect spherical
or isotropic expansion is indicated by a small but definitive tilt in
the line profiles. This tilt runs from $-$29.73~\kms\, in the south
(bottom of the profile) to $-$12.07~\kms\, (top) in the north. This
must be produced by a slightly anisotropic expansion of the bubble,
i.e.  receding velocities in the northern part dominate over the
approaching ones in the southern section of the bubble, with respect
to the systemic velocity. If this trend continues it is then expected
that Kn~61 will become an elongated nebula, tending to an elliptical
shape with time.  A crude estimate of the kinematic age for the nebula
is calculated considering that the angular diameter of the nebula is
104\arcsec. For an assumed distance of 4~kpc, its linear radius is
0.96~pc. Considering as constant the expansion velocity quoted above,
this yields a kinematic age for the bubble of 1.6$\times$10$^4$~yrs.

Likewise, the medium resolution nebular spectrum for Kn~61 (Figure~5,
righthand panels) shows only \OIII\,4959,5007~\AA, \Ha\, and very weak
HeII\,4686~\AA\, and H$\beta$ nebular emission lines.  The sky lines
have been removed.  The nebular medium-resolution spectrum from
the SAO-RAS (Figure~7, right panel) confirms the previous results for
the corresponding wavelength ranges.  A list of the main emission
lines and their fluxes are presented in Table~3. The nebular spectrum
of Kn~61 does not show measurable [O~III]\,4363~\AA\, neither the [N~II]
lines. No other diagnostic line ratios are available within this
wavelength range, therefore it is not possible to derive electron
temperature nor electron density and consequently ionic abundances.
It is interesting to notice however that the low H$\alpha$ / H$\beta$
flux ratio measured seems to indicate that Kn 61 is at least partially
optically thin to Lyman photons (e.g. Osterbrock, 1974) which would
contribute to explain the very faint H$\alpha$ emissivity of this
nebula.

\subsubsection{Central star of Kn~61}

Figure~5 (lefthand panels) shows the spectra from OAN-SPM of the
central star of Kn~61.  The spectrum has been binned $\times$4 in the
spectral axis to improve signal to noise and help highlight
photospheric lines.  The stellar spectra show lines characteristic of
a Group 3 = lg E (low gravity central star) class of PG~1159 star
\citep[e.g.][]{Par98}, revealed by the absorption line at
C~IV\,4647~\AA\, and the double emission lines also from
C~IV\,5801,5812~\AA. These lines are more clearly appreciated in the
medium resolution stellar spectrum of Kn~61 obtained at SAO-RAS (see
Figure~7, left panel). The later spectrum for the central star
indicates the presence of an emission line at 4713~\AA\, this is
likely a nebular contamination from either He~I or [Ar~IV], as it is
sometimes observed in other PG~1159 stars \citep{Par98}. Additionally
the C~IV\,4647~\AA\, absorption line is detected. The PG~1159 stars
are considered to be very hot, hydrogen deficient post-AGB stars. In a
recent survey of such stars detected in the Data Release 10 of the
Sloan Digital Sky Survey \citep{2014A&A...564A..53W} seven stars are
of PG1159-type with spectra similar to the central star of
Kn~61. Particular close resemblance is found between the spectra of
the central star of Kn~61 and SDSS J075415.12+085232.18, a PG1159 star with
120\,000\,K, log\,$g=0.33$ and C/He=0.33, where the C~IV\,4647
\AA\ has also been identified.

 We have monitored the photometric behavior of the central star of
 Kn~61 for 2-3 hours each night for 16 consecutive nights.  No strong
 amplitude or periodic variability beyond the natural data scatter on
 the order of 0.028 mag was detected on such short time scales.
 Therefore, the photometric data were averaged for each night and the
 resulting light curve covering 16 nights was analyzed for the
 presence of a periodic signal using the Period04 program.  The latter
 employs a discrete Fourier transformation technique to calculate the
 power spectra and spectral window corresponding to the analyzed
 time-series.  This produced a statistically significant peak at a
 frequency of 0.176 days$^{-1}$ in the power spectrum (see Figure 8
 left panel), which corresponds to a period of 5.7($\pm$0.4)
 days. The semi-amplitude of variability is A$\approx 0.02$\,mag and
 is comparable with the root mean square (RMS) of individual nightly
 light curves before averaging. Similar analysis on two field stars of
 comparable brightness proved that the periodic signal is authentic
 and far exceeds random noise.  The significance level corresponding
 to 99\% is marked in Figure 8 by a horizontal dashed line. In
 addition we frequency folded the power spectrum within the spectral
 window, which helps to sort out the alias periods produced by uneven
 time series. The result confirms a strong, single peak at the
 indicated frequency. The light curve folded with the determined
 period is presented in the right panel of Figure 8. The best fit
 $sin$ curve with the A=0.0176 amplitude around the average magnitude
 m$_{\mathrm R}=18.674$ is also plotted.  The phase zero is set
 arbitrarily to the moment when the object has an average
 brightness. The error bars in Figure 8 (right panel) reflect the RMS
 of individual points around each night's average and they far exceed
 statistical errors of aperture photometry.

\citet{Long13} analyzed the light curve of the central star of Kn~61 and
found cyclical variability which might be periodic on time scales from
2 to 6 days.  Our result does not contradict their non-conclusive
findings. The object requires longer monitoring and more uniform
observations to determine the origin of the variability. 


It is interesting to notice, however, that for more than a couple of
decades binary nuclei in PNe have been suggested as responsible for
shaping non-spherical, axis-symmetric or point-symmetric nebulae
(e.g. Livio \& Soker 1988) and as a source of their collimated
outflows. Kn~61 does not display collimated outflows and it is close
to spherical symmetry.  However, if the central star of Kn 61 is a
binary, its possible orbital period of $\approx$ 6 days is relatively
long compared to other binary CSPN. The majority of close binary CSPN
have usually periods of only a fraction of a day \citep{deMarco09}.  A
longer period means that the secondary is relatively distant, leading
to reduced chance of interaction between stars and thus reduced
influence on the nebular shape and collimation of an outflow.

It is also of interest to compare Kn~61 with the PNe NGC~7094 and
Abell~43 (Rauch et al. 2005).  These PNe have a very similar morphological structure to
Kn~61, i.e. they are both roughly spherical filamentary bubbles
and in both cases their central stars are classified as hybrid-PG~1159
type (Solheim et al. 2007). They also display high expansion velocities, $\geq 50$ \kms.
These stars are thought to have
experienced a late thermal pulse while they were still on the
asymptotic giant branch and hydrogen-shell burning was still on, 
  yet they are able to keep small amounts of atmospheric hydrogen.
We have drawn the images and spectra of these PNe from the {\it ``The
  San Pedro M\'artir Kinematic Catalogue of Galactic Planetary
  Nebulae''} (L\'opez et al.  2012) and these are reproduced in
Figure~9.  Notice the remarkable similarity with Kn~61, both in the
images and the bi-dimensional line profiles. In these cases the \Ha\,
line is bright, in contrast with Kn~61. Also in both cases the \NII\,
lines are absent, as in Kn~61 (see L\'opez et al. 2012). Pereyra et
al. (2013) found from an analysis of 100 evolved PNe that those
objects located at the point of maximum temperature in the
evolutionary tracks, right before their luminosity starts dropping
towards the white dwarf region, exhibit a lack of \NII\ emission. 
The lack of [N~II] is a consequence of the small fraction of nitrogen in the
singly ionized state. This, in turn, follows from the high
ionization levels of the nebula due to a very hot central star that
photoionizes most of the N into doubly and triply ionized states.
The relation between soap bubbles shapes, i.e. highly filamentary
spherical bubbles, and the PG~1159 type stars is puzzling and is
pointed out here for the first time.

\subsection{Pa 5.}

Figure~2 shows the corresponding images for Pa~5. We did not detect
any emission in \NII\,6584~\AA\, nor in \SII\,6717,6731~\AA\, in our
images. The \OIII\, and H$\alpha$ images show a bright bar that looks
like an equatorial structure with a faint lobe extending perpendicular
to it towards the northeast.  A much fainter extension is barely
appreciated on the opposite side.  The slit was located along the
bright structure. The P--V arrays (see Figure~4) from this position
confirm that this structure has an expanding doughtnut-like or
toroidal shape, expanding at 45.2($\pm$2)~\kms\, and with a systemic
velocity of 12.4($\pm$2)~\kms. The \Ha\, line profile looks filled
close to the top and bottom cusps indicating that the toroid has a
thick inner wall (see Figure~2, right panel).  Unfortunately weather
conditions precluded us from obtaining a second slit position along
the polar extensions that would have help characterized the likely
bipolar structure, as suggested by Kronberger et al. (2006). In a
study of PNe with close binary nuclei L\'opez et al. (2011) find that
the large majority of these objects tend to have equatorial density
enhancements likely produced by the ejection of the envelope along the
equatorial plane after the common envelope phase. If the nucleus of
Pa~5 is confirmed as a close binary system (see below) its morphology
would agree with this characteristic.

The one-dimensional medium-resolution nebular spectra of Pa~5 are
shown in Figure~6 (panels b \& d). The emission lines \Ha,
[O~III]\,5007,4959~\AA, \Ha, H$\beta$, H$\gamma$,
He~II\,4686,5412~\AA\, and [Ar~IV]\,4740,4711~\AA\, stand out in the
spectra. A list of the main emission lines and their fluxes are
presented in Table~3. As for Kn 61 the [O III]\,4363~\AA\, line is too
weak to measure and the [N II] lines are absent, it is therefore not
possible to derive a relieble electron temperature in the available
wavelength range.  For the case of the electron density, the [Ar IV]
lines are present and their line ratio indicates an electron density
of the order of 1100($\pm$200) cm$^{-3}$, assuming an
electron temperature of 10$^4$ K. This relative high density is
reasonably consistent with the fact that the spectrum comes from the
dense waist or equator of this PN.

\subsubsection{Central star of Pa~5}

 {\O}stensen, et al. (2010) using low resolution spectroscopy in the
 spectral range of 3500~--~5300~\AA, found that the central star of
 Pa~5, J19195$+$4445, is a very hot star and they identify it as
 belonging to the PG~1159 on the basis of the presence of C~IV and
 He~II close to 4686 \AA. The spectra for the central star of Pa~5,
 shown in Figure~6 (panel a \& c) reveal the presence of lines from
 the Pickering series He~II \,4200, 4541, 4686, 5412~\AA; H$\beta$,
 H$\gamma$ and Ca II\, 5889~\AA, but we do not detect the C IV lines so
 we are unable to confirm the PG~1159 nature of this star.

In addition, {\O}stensen et al. (2010) using the Kepler data find a
period of 1.12 days for the central star of Pa 5. They report a very
small amplitude of only 0.05\% in terms of the significance of the
variability and consider the origin of the variability as
unknown. Although we attempted to measure radial velocity and
photometric variations over several nights, the period of this object
being close to one day precludes observing different phases from a
single ground-based telescope and we could not detect any orbital
modulation.

\section{Conclusions}
\label{sec:conclusion}

Kn~61 is a higly filamentary bubble with a high expansion velocity,
67.6~\kms.  Kn~61 does not show emission in \NII, and \SII\, due to a
very hot central star that photoionizes most of the N into doubly and
triply ionized states.  The H$\alpha$ emission is extremely faint in
comparison to \OIII, showing some limb brightening at the border of
the nebula.  The nebular and stellar emission of Kn~61 suggests that
the central star had nearly exhausted its hydrogen layers as it left
the AGB.  From medium-resolution spectra of the central star, we
detected the absorption line at C~IV\,4647 \AA, the emission line
C~IV\,4659 \AA, and the double emission lines also from
C~IV\,5801,5812~\AA, which are characteristic of PG~1159 hydrogen
deficient, post-AGB stars.  The morphology of Kn~61 is very similar to
NGC~7094 and Abell~43, whose nebular spectrum has the same kinematic
behavior as that of Kn~61.  None of these nebulae present \NII\,
emission lines, though the \Ha\, line emission is bright in NGC~7094
and Abell~43, in contrast to Kn~61.  NGC~7094 and Abell~43 have
central stars of hybrid-PG~1159 type, they both are pulsators
\citep{Solheim07}, they present He~II/C~IV\,4650-4686~\AA\, and and
more prominent Balmer lines in the associated nebula. These objects,
including Kn~61, are likely mature nebulae located in the region of
maximum temperature in their corresponding evolutionary tracks in the
Hertzsprung Russell diagram, right before their luminosity starts
dropping towards the white dwarf region. The lack of \NII\ emission is
a characteristic of PNe in this stage of evolution, since low
ionization states are pumped to higher levels and is also where the
largest expansion velocities are observed, as pointed out by Pereyra
et al. (2013) . The possible link between filamentary bubbles and
PG~1159-type stars is noted for the first time and requires additional
analysis.  From the CCD photometry performed on the central star of
Kn~61 we obtained a light curve with an orbital period for the system
of $\sim$5.7($\pm$0.4) days, this coincides with preliminary Kepler
data presented by Long et al. (2013) and indicates the possible
presence of a binary core, though longer monitoring and uniform
observations are required to exclude other explanations, such as
variations driven by pulsations.  Although Kn 61 is close to having a
spherical outline or shape, a deviation from spherical symmetry and
isotropic expansion has been detected in the long-slit spectra. A
crude kinematic age for the bubble yields 1.6$\times$10$^4$~yrs.

Pa~5 does not show \NII, and \SII\, emission lines either.  Its
morphology is dominated by a dense equatorial toroid and faint polar
extensions. The \OIII\, and \Ha\, P--V arrays show that equatorial
enhancement is thick, sitting at a systemic velocity of
12.4($\pm$2)~\kms\, and expanding at 45.2($\pm$2)~\kms.  The central
star is very hot, showing prominent
He~II\,4541,4686,5412 \AA\,, H$\beta$ and
Ca~II\,5889 \AA\, absorption lines. Kepler data have yielded a period
of 1.12 days for the central star with a very small amplitude
\citep{Ostensen10} the origin of this variability remains unknown. Our
CCD photometry and spectroscopy could not confirm any orbital
modulation in this case.

\acknowledgments

Support for this study was provided by CONACyT, M\'exico through a
research studentship to D.G.B.  This research has benefited from the
financial support of PAPIIT-UNAM through project IB100613-RR160613.
This research is based upon observations acquired at the Observatorio
Astron\'omico Nacional in the Sierra San Pedro M\'artir (OAN-SPM),
Baja California, M\'exico and from the Special Astrophysical
Observatory 6-m telescope of the Russian Academy of Sciences.  We are
grateful to the excellent support of the technical personnel at the
OAN-SPM. We thank the anonymous referee for the constructive comments
that improved the presentation of this work.

\begin{table*}[!t]
\centering
\caption{Log  of time-resolved  observations of {\bf Kn61}.}
\begin{tabular}{l|cccc} \hline 
&  \multicolumn{4}{c}{Spectroscopy medium resolution/ 2.1m  OAN-SPM}            \\[0.1pt]
\hline 
Date  & No. of   &   Exp. time & Range &  Resolution             \\
DD/MM/20YY   &     frame         &    (s)   &  \AA    &    \AA      \\[1pt]   \hline 
 11/08/13        &  1  & 1200 &  4000 - 7000 & 4.2 \\
 12/09/12        &  2 & 1200 & 4560- 5750 & 2.0 \\
 17/07/12        &  3 & 1200 & 4000- 7000 & 4.2\\  
 16/07/12        &  1 & 1200 & 4560- 5750 & 2.0 \\
 09/06/12        &  6  & 1200 & 4560- 5750 & 2.0 \\ 
\hline
\hline 
&\multicolumn{4}{c}{Spectroscopy medium resolution / 6.0m  SAO-RAS }              \\[1pt]
\hline 
12/09/13 & 3 & 600 & 3900 - 5700& 5\\
\hline
\hline 
&\multicolumn{4}{c}{Spectroscopy hight resolution / 2.1m OAN-SPM}              \\[1pt]
\hline 
25/05/12 & 1 & 1800 & H$\alpha$, \OIII & 0.26 (H$\alpha$)\\
  & &  & & 0.19 (\OIII)\\
\hline\hline
&\multicolumn{4}{c}{Photometry / 1.5m OAN-SPM }              \\ \hline
Date  & No. of   &   Exp. time & filter  &               \\
DD/MM/20YY   &     frame         &    (s)   &              \\[1pt]   \hline 
4,5,6,8,13/07/13      &  19  &  90   &\sc{r}     \\
14,16,17,31/07/13   &  19  &  90   &\sc{r}      \\
1,2,4,10/08/13         & 20   &  90   & \sc{r}            \\
11,13,28/08/13        & 20   &  90   & \sc{r}            \\
\hline
\hline 
&\multicolumn{4}{c}{CCD narrowband images / 0.84m OAN-SPM }              \\ \hline
 03/08/12 &  1 & 1800 & H$\alpha$, \OIII  \\
\hline
\hline
\end{tabular}
\label{table:kn61}
\end{table*}

\begin{table*}[!t]
\centering
\caption{Log  of time-resolved  observations of {\bf Pa5}.}
\begin{tabular}{l|cccc} \hline 
&  \multicolumn{4}{c}{Spectroscopy medium resolution/ 2.1m  OAN-SPM}            \\[0.1pt]
\hline 
Date  & No. of   &   Exp. time & Range &  Resolution             \\
DD/MM/20YY   &     frame         &    (s)   &  \AA    &    \AA      \\[1pt]   \hline 
16/07/12       & 1  & 1200 &  4000 - 7000& 4.2\\
18,20,24/07/12        & 17  & 1200 &    4250 - 5500 & 2.0\\ 
12,14,15,16/09/12       & 17 &  1200 &   4550 - 5750 & 2.0 \\ 
10,12/08/13      & 3  & 1200 &   4000 - 7000 & 4.2\\ 
11/08/13      & 1  & 1200 &    4700 - 5700 & 2.0\\
13,14,15/08/13     &  22  &  1200 &  4450 - 5600 & 2.0\\
\hline
\hline 
&\multicolumn{4}{c}{Spectroscopy hight resolution / 2.1m OAN-SPM}              \\[1pt]
\hline 
25/05/12 & 1 & 1800 & H$\alpha$, \OIII & 0.26 (H$\alpha$)\\
  & &  & & 0.19 (\OIII)\\
\hline\hline
&\multicolumn{4}{c}{Photometry / 1.5m OAN-SPM }              \\ \hline
Date  & No. of   &   Exp. time & filter  &               \\
DD/MM/20YY   &     frame         &    (s)   &              \\[1pt]   \hline 
5,6,8,13,14/07 & 20/20 & 90/90 & \sc{r/i} \\ 
16,17,28,30,31/07 & 20/20 & 90/90 & \sc{r/i}\\ 
1,2,4,10/08    & 20/20 & 90/90 & \sc{r/i}  \\
11,13,14,28/08   & 20/20 & 90/90 & \sc{r/i}  \\
31/08   & 20 & 90 & \sc{i}  \\
\hline
\hline 
&\multicolumn{4}{c}{CCD narrowband images / 0.84m OAN-SPM }              \\ \hline
 09/08/13 &  1 & 1800 & H$\alpha$, \OIII  \\
\hline
\hline
\end{tabular}
\label{table:pa5}
\end{table*}

\begin{table*}
 \centering
    \caption{Fluxes of the main emission lines }
\begin{tabular}{lccccc  |  lcccc} \hline
{\bf Kn~61}&&&&& &{\bf Pa~5}&&&\\\hline
ID  & Wavelength Rest    & F$_{\lambda}$/F(H$\beta$)   &   &    &  & ID  & Wavelength Rest      & F$_{\lambda}$/F(H$\beta$)   &     \\
    &  \AA         &    measured   &           &          &       &   &    \AA     &    measured  & \\
            \hline
  He  {\sc ii}     & 4685.68	&  0.88    &	       &  &  & H  {\sc i}   & 4340.47  	&  0.53     &	      \\
   O {\sc iii}         &  4958.93   &  4.27    &	   &    &	&	 He {\sc ii}	 &  4685.70               &   1.27 &  &	   \\
O {\sc iii} & 5006.85 & 14.42 &  & & &  Ar   {\sc iv}        & 4711.33	&  0.24    &	   \\   
 
  H  {\sc i}     & 6562.82	&  2.69     &	   &    && Ar {\sc iv}  &	4740.19	 	&  0.18     &	  \\
&&&  &    && O {\sc iii}  	 &       4958.93        &   1.35   &                   \\	

	           	&&&&	 &               &   O {\sc iii}   &       5006.85	    &   4.16 &    \\	      
	           	&&&&	 &               &   He {\sc ii}   &       5411.57	    &  0.14 &    \\
	           	&&&&	 &               &   H {\sc i}   &       6562.82	    &   2.94   &  \\	      	     
	           	&&&&	 &               &     &      	    &   &    \\	      	    
logF(H$\beta$) &  \multicolumn{3}{c}{-14.49 erg cm$^{-2}$s$^{-1}$}   &         &        &                     logF(H$\beta$)  &              \multicolumn{3}{c}{-13.66 erg cm$^{-2}$s$^{-1}$}    &     \\

\hline \hline
\end{tabular}
\label{tab:emlines}
\end{table*}

\begin{figure*}[t!]
\begin{center} 
   \includegraphics[width=1\textwidth]{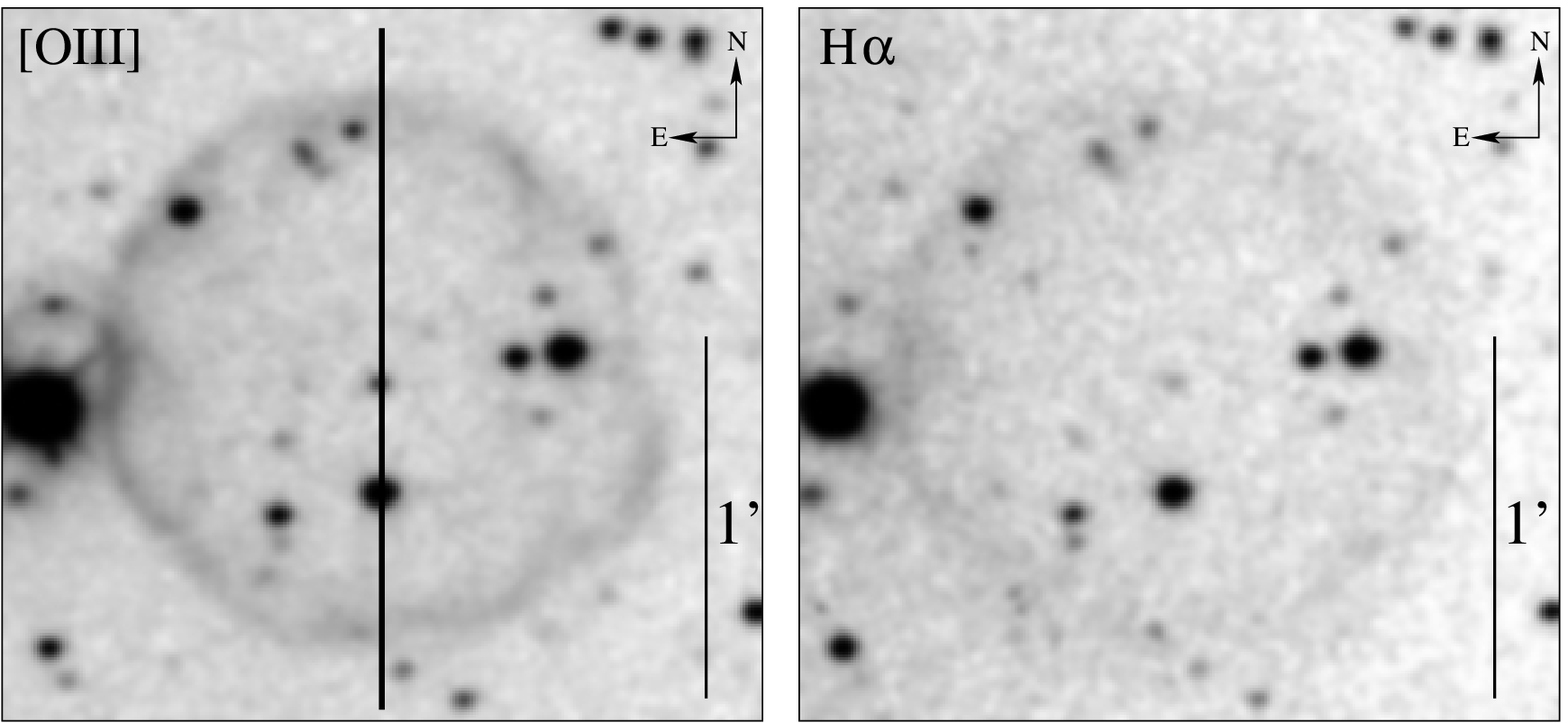}
\end{center} 
   \caption{Images of Kn~61 obtained at the OAN-SPM 84~cm telescope,
     in the light of \OIIIlam{} (left panel) and H$\alpha$ (right
     panel). Location of the long-slit position is indicated on the
     \OIII{} image. }
  \label{fig:Figure1}
\end{figure*}

\begin{figure*}[t!]
\begin{center} 
    \includegraphics[width=1\textwidth]{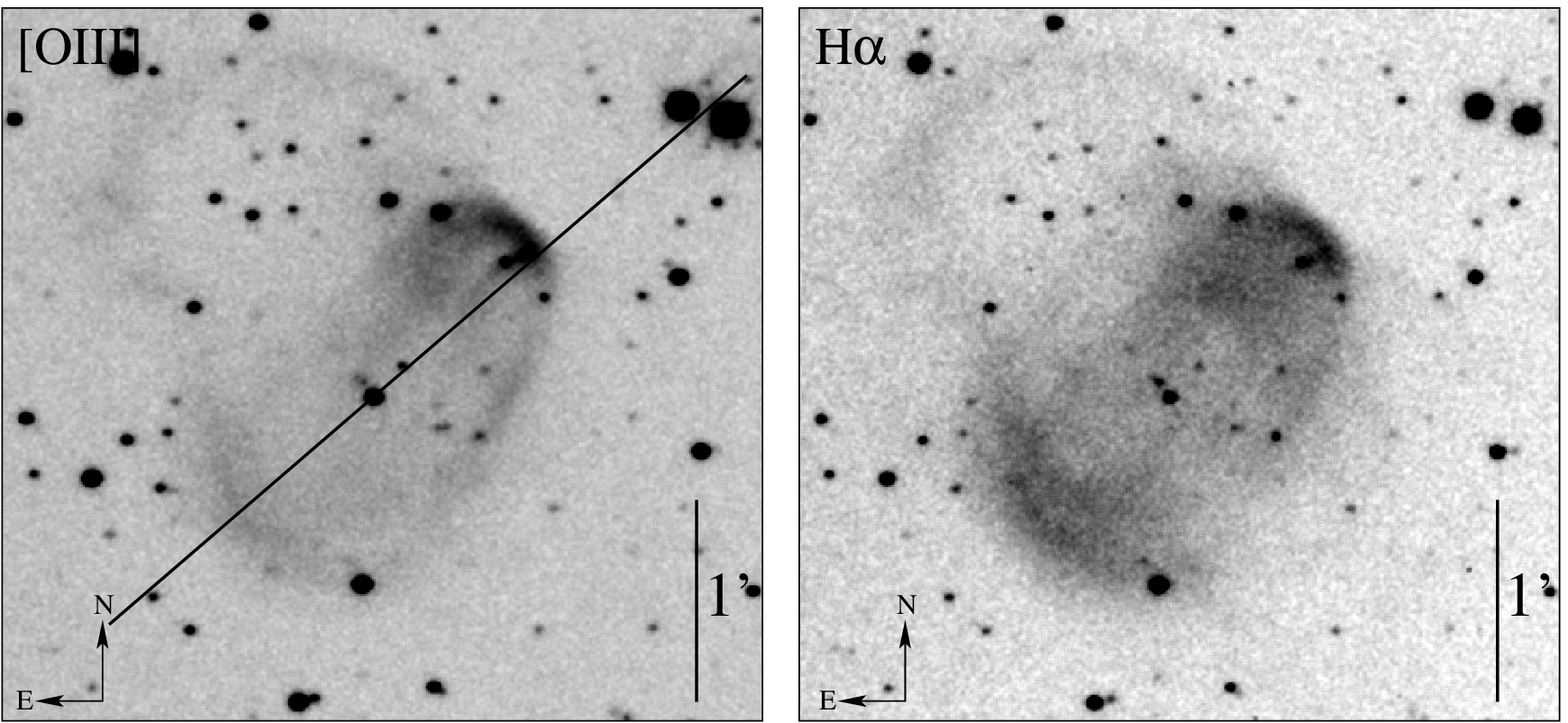}
\end{center} 
    \caption{Images of Pa~5 obtained at the OAN-SPM 84~cm telescope in
      the light of \OIIIlam{} (left panel) and H$\alpha$ (right
      panel). Location of the long-slit is indicated on the \OIII{}
      image }
  \label{fig:Figure2}
\end{figure*}

\begin{figure}[t!]
  \centering
    \includegraphics[width=0.7\textwidth]{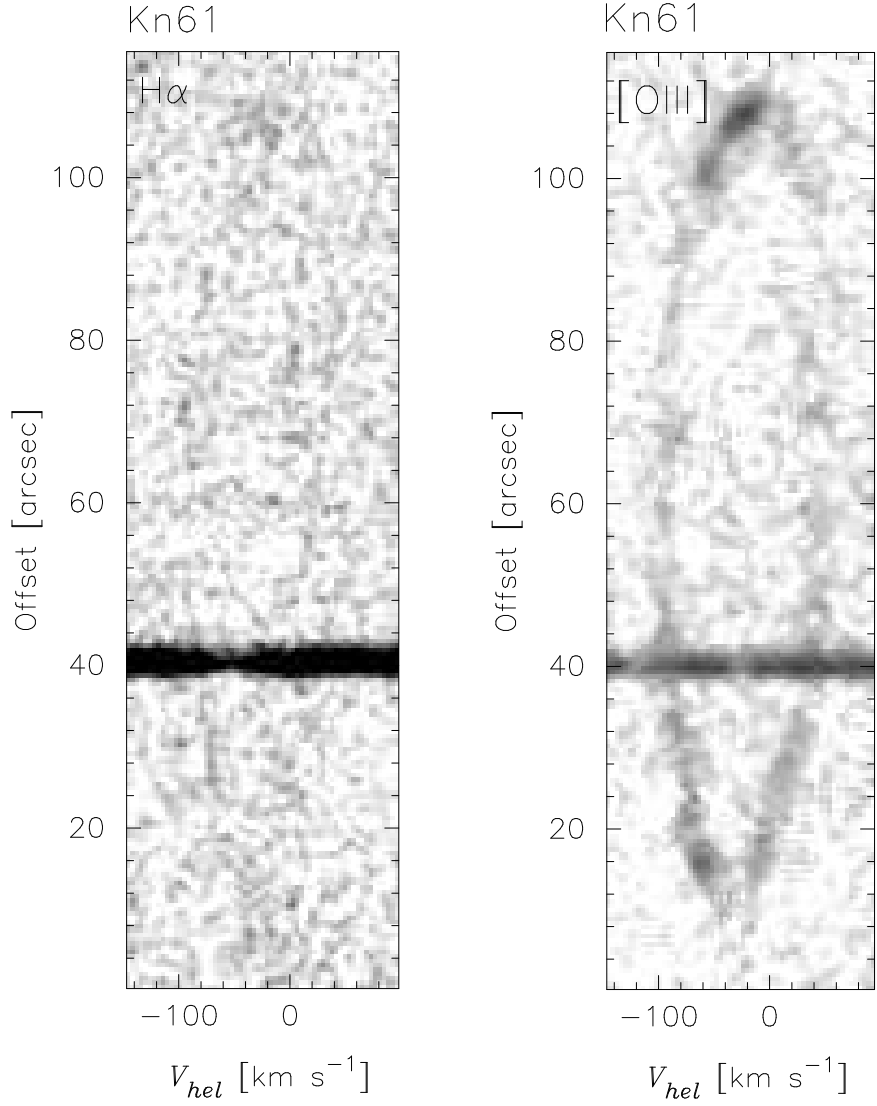}
  \caption{Kn~61 Bi-dimensional P--V array, 
    the continuum from a field star has not been subtracted}
  \label{fig:Figure3}
\end{figure}

\begin{figure}[t!]
    \centering
    \includegraphics[width=0.7\textwidth]{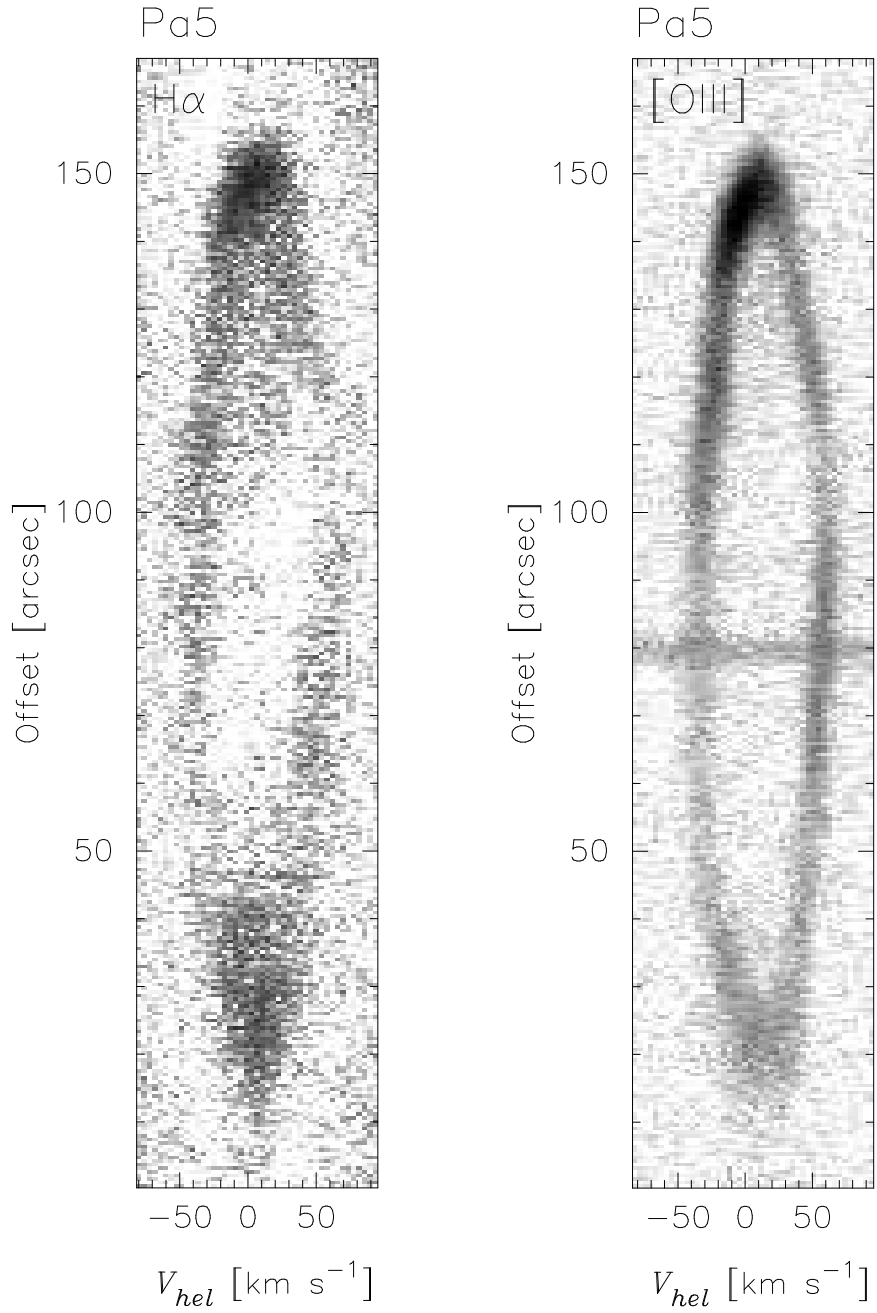}
  \caption{Pa~5 Bi-dimensional P--V array. The continuum from the central
  star has been subtracted  from the H$\alpha$ P --V array.}
  \label{fig:figure4}
\end{figure}

\begin{figure*}[t!]
\centering
 \includegraphics[width=1\textwidth]{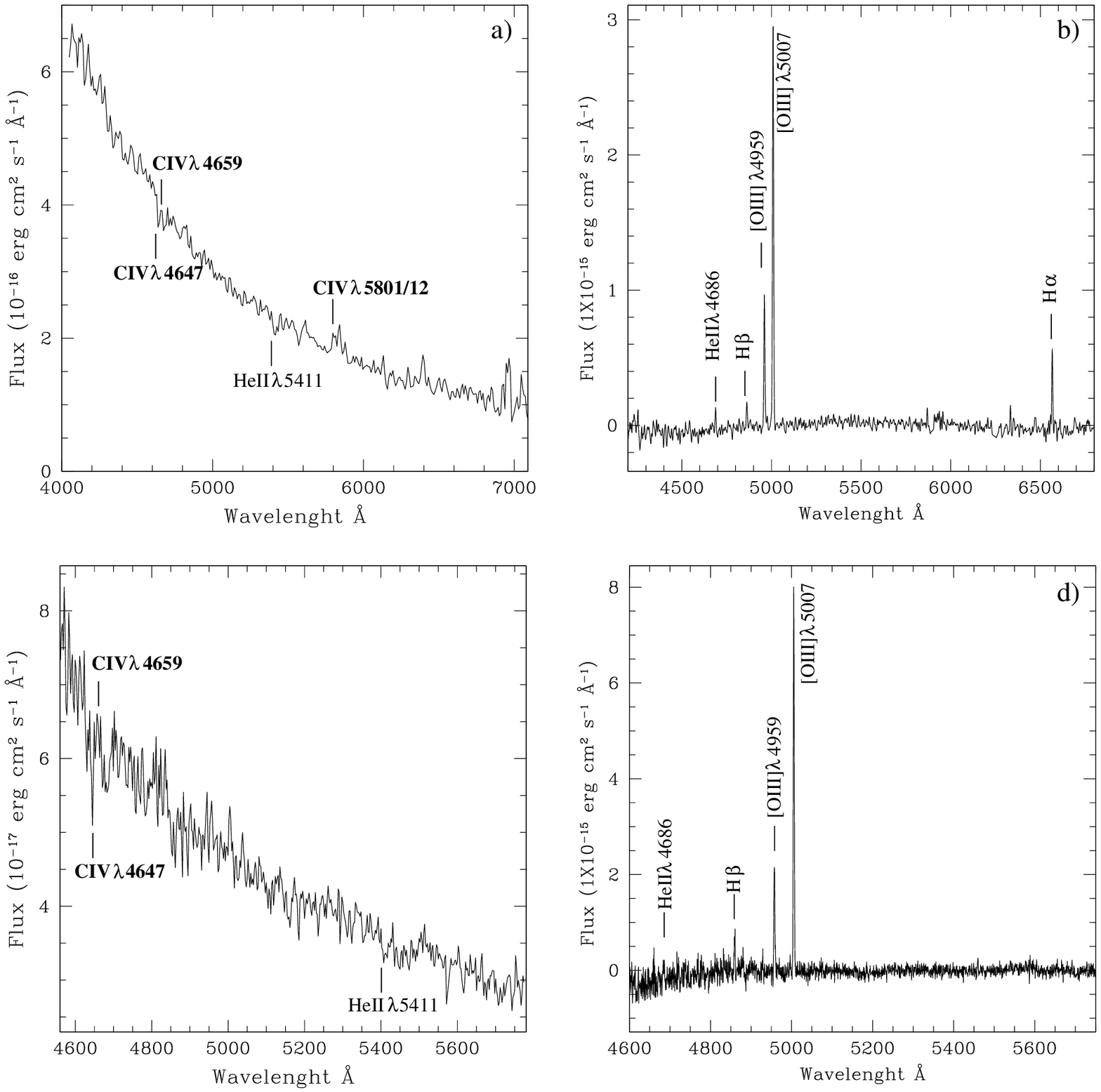}
  \caption{Panels a) and b): OAN-SPM one-dimensional stellar and
    nebular spectra of Kn~61, respectively, taken with a 600 l
    mm$^{-1}$ grating. Panels c) and d): OAN-SPM one-dimensional
    stellar and nebular spectra of Kn~61, respectively, taken with the
    1200 l mm$^{-1}$ grating.}
  \label{fig:figure5}
\end{figure*}

\begin{figure*}[t!]
\begin{center}
 \includegraphics[width=1\textwidth]{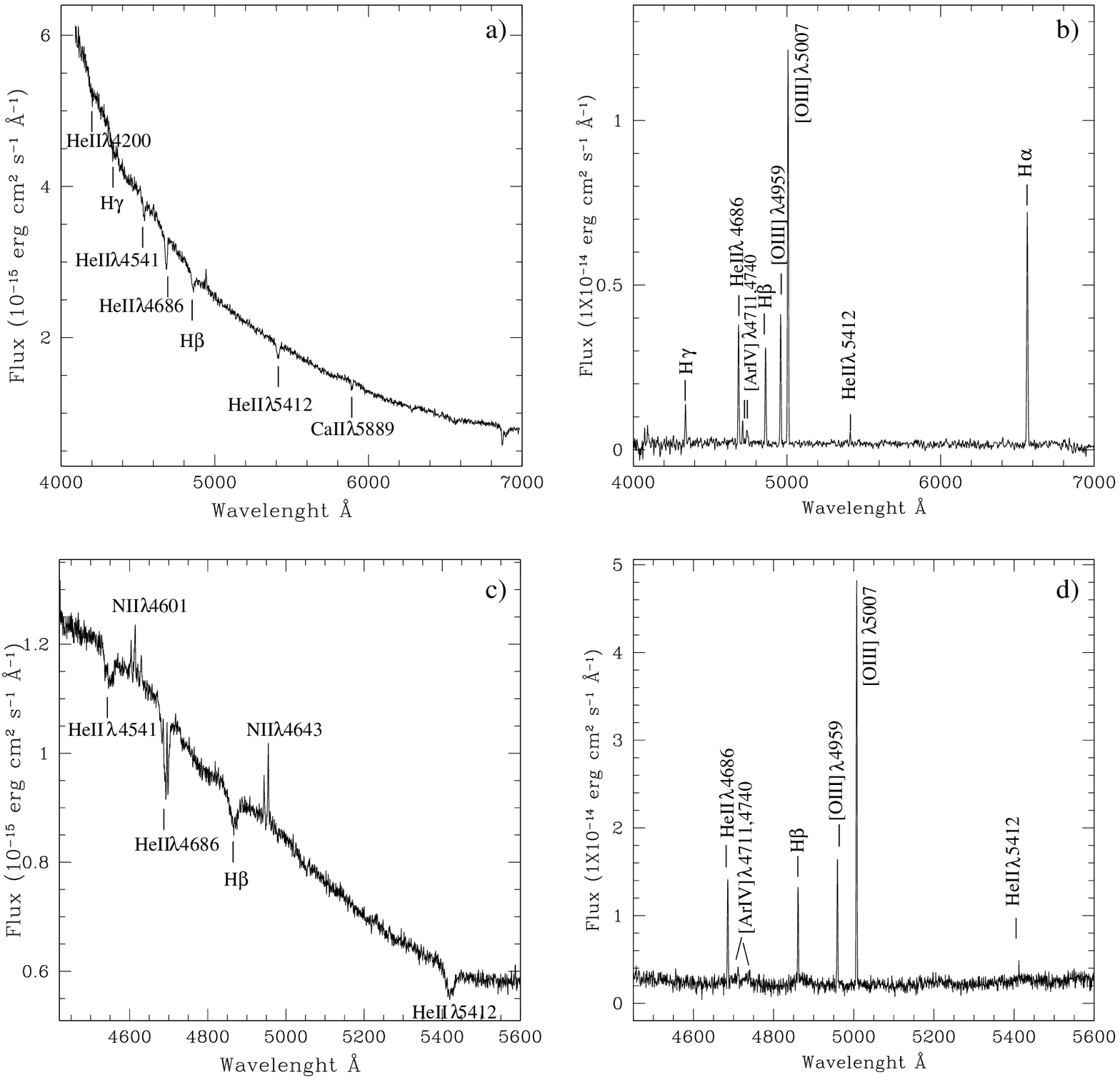}
\end{center}
  \caption{Panels a) and b): OAN-SPM one-dimensional stellar and
    nebular spectra of Pa~5, respectively, taken with a 600 l
    mm$^{-1}$ grating. Panels c) and d): OAN-SPM one-dimensional
    stellar and nebular spectra of Pa~5, respectively, taken with the
    1200 l mm$^{-1}$ grating.}
  \label{fig:figure6}
\end{figure*}

\begin{figure*}[!t]
\begin{center}
  \includegraphics[width=1\textwidth]{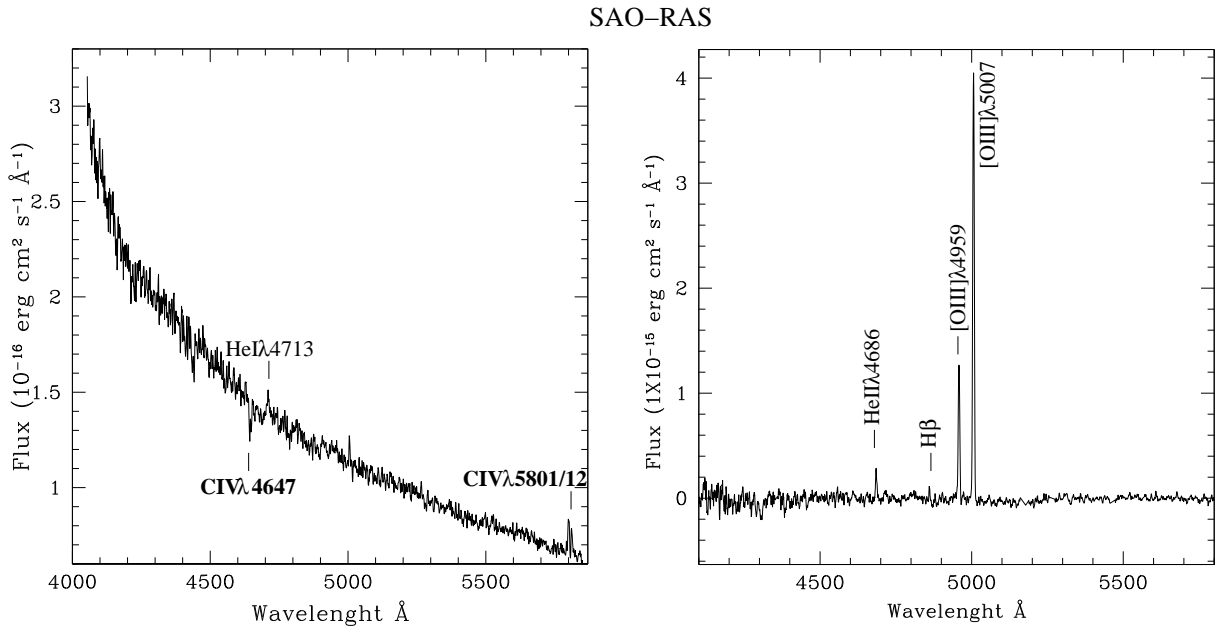}
\end{center}
  \caption{SAO-RAS spectra. {\it Left panel)} Stellar spectra of SDSS
    J192138.93+381857.2. {\it Right Panel)} Nebular spectrum of Kn~61}
  \label{fig:figure7}
\end{figure*}

\begin{figure*}[!t]
\begin{center} 
   \includegraphics[width=1.0\textwidth]{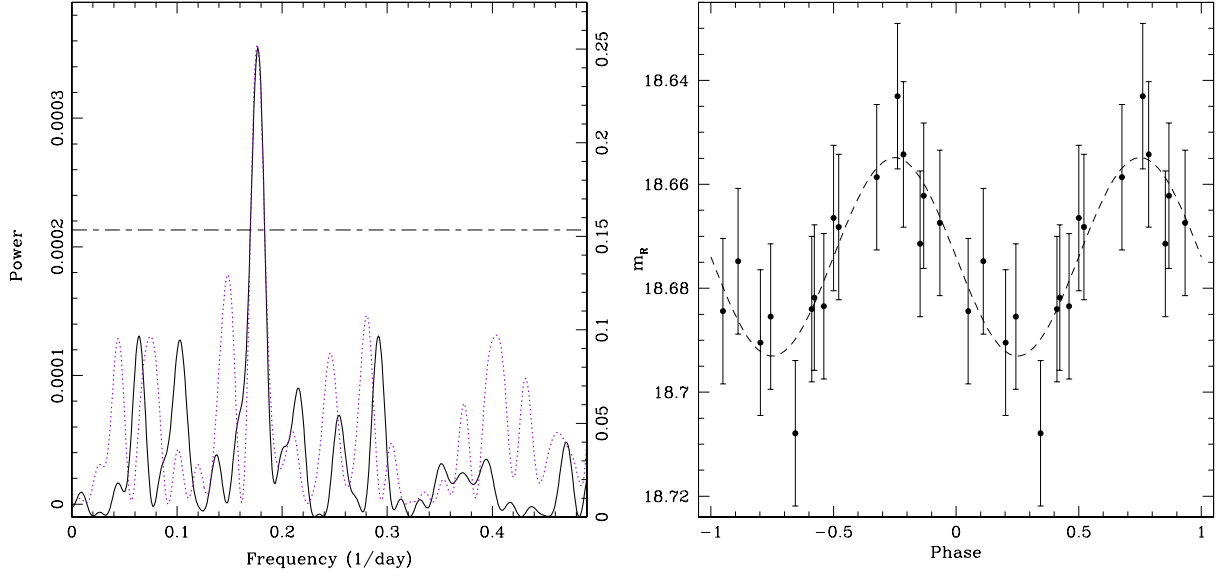}
\end{center} 
   \caption{ {\it left panel:} Power spectrum of  the photometric light curve. 
   The dotted line represents the raw power spectrum calculated by using average nightly brightness and its power scale is indicated by the left vertical axis. The solid line is the power spectrum obtained by {\it CLEAN}ing alias signals using the spectral window.  Both curves are scaled to match the peak power, but the signal in the {\it CLEAN}ed spectrum is different, and its scale is indicated on the right-side axis. 
    The power  peaks at the  0.176$d^{-1}$ frequency. The horizontal dashed line indicates 99\% confidence level of the raw power spectrum.  
       {\it right panel:} The light curve of the  Kn~61 central star  folded with the 5.7 day period.  The error bars correspond to the RMS of individual night light curves, the points correspond to the nightly brightness average. The best-fit $sin$ curve is overlaid on the data.}
  \label{fig:figure8}
\end{figure*}

\begin{figure}[!h]
\begin{center} 
   \includegraphics[width=0.8\textwidth]{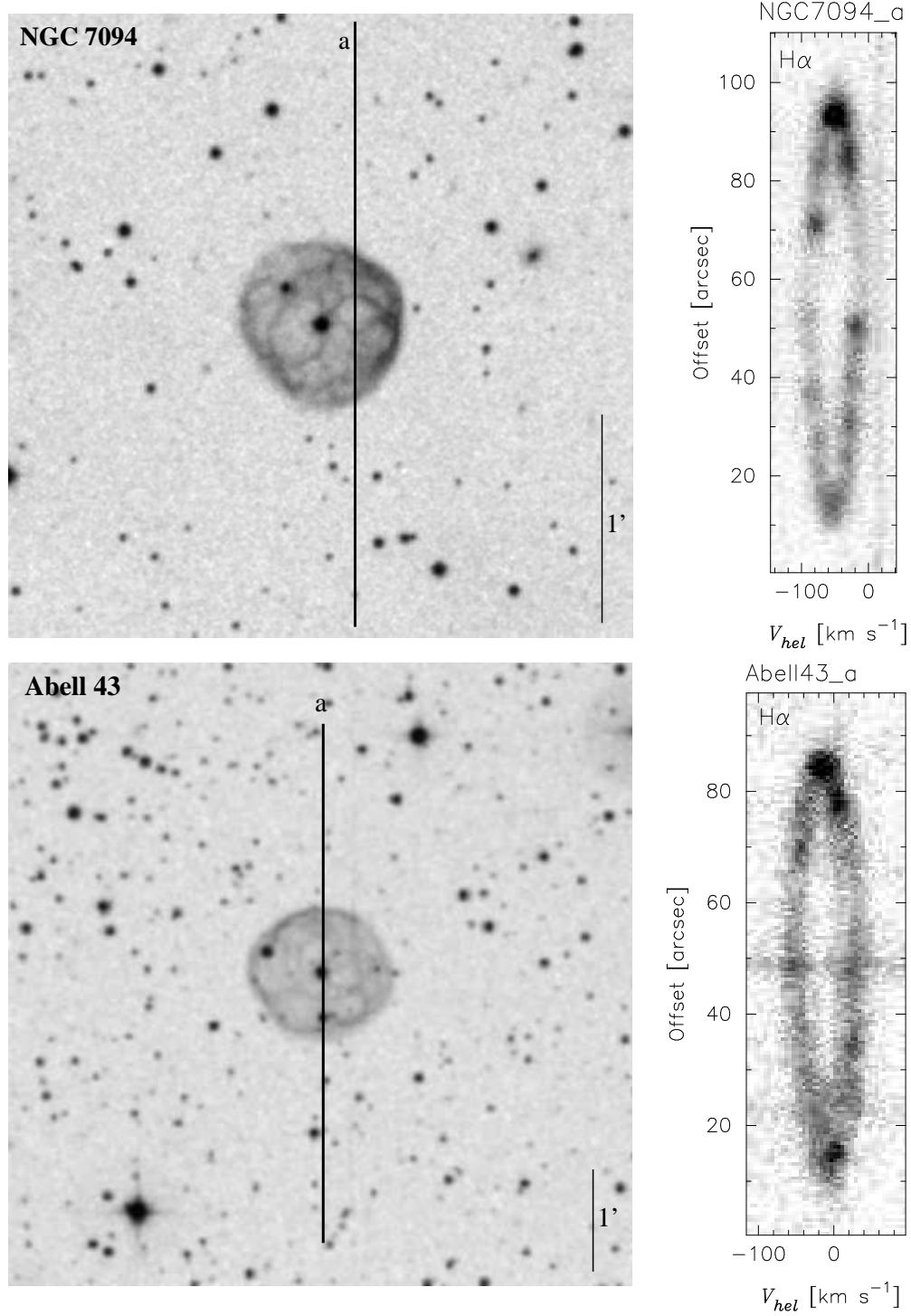}
\end{center} 
   \caption{Images and long-slit spectra of NGC 7094 and Abell 43
     taken from the SPM Kinematic Catalogue of Galactic Planetary
     Nebulae, http://kincatpn.astrosen.unam.mx/. Notice their
     similarity in both, images and spectra, with Kn 61}
   \label{fig:figure9}
\end{figure}

\end{document}